\documentclass[epj]{webofc}
\usepackage[varg]{txfonts}   
\usepackage{subfigure}
\usepackage{graphicx}
\usepackage{hyperref}
\newcommand{\be}{\begin{equation}}
\newcommand{\ee}{\end{equation}}
\newcommand{\bea}{\begin{eqnarray}}
\newcommand{\eea}{\end{eqnarray}}
\newcommand{\nn}{\nonumber}

\begin{document}
\title{Strong Sector in non-minimal SUSY model}
%
%

\author{\firstname{Antonio} \lastname{Costantini}\inst{1}\fnsep\thanks{\email{antonio.costantini@le.infn.it}}}

\institute{Dipartimento di Matematica e Fisica "Ennio De Giorgi", \\ Universit\`a del Salento and INFN-Lecce, \\ Via Arnesano, 73100 Lecce, Italy}

\abstract{We investigate the squark sector of a supersymmetric theory with an extended Higgs sector. We give the mass matrices of stop and sbottom, comparing the Minimal Supersymmetric Standard Model (MSSM) case and the non-minimal case. We discuss the impact of the extra superfields on the decay channels of the stop searched at the LHC.}
\maketitle
\section{Introduction}
The search of supersymmetric partners of the known particles, expecially the quarks, is one of the main purpose of the LHC. They are supposed to be heavy because supersymmerty (SUSY) is broken, otherwise they would be observed at the electroweak symmetry breaking (EWSB) scale. Since the start of LHC with Run 1, the lower bound on the mass of the lightest squark, which is usually taken to be the lightest stop, has increased \cite{stopmass}. In this paper we are going to show how much these bounds are related to the assumptions that the supersymmetric model is minimal in its superfield content. In section \ref{mssm} we briefly review the squark sector in the MSSM, giving the expression of the mass matrices for stop and sbottom. In section \ref{tnmssm} we show how the squark sector changes in the case of non-minimal SUSY and in section \ref{branch} we present the results for the branching ratios of the lightest stop decay channels tested at the LHC. Section \ref{concl} is left to the conclusions.  
\section{Squark Mass Matrices in MSSM}\label{mssm}
As it is well known, in the minimal version of SUSY each particle of the Standard Model (SM) has a superpartner and they are related each other by a SUSY transformation. The superpotential of the MSSM is
\begin{equation}
\mathcal{W}=\mathcal{W}_{MSSM}+\mathcal{W}_{H}
\end{equation}
with
\begin{equation}
\mathcal{W}_{MSSM}= y_t \hat U \hat H_u\!\cdot\! \hat Q - y_b \hat D \hat H_d\!\cdot\! \hat Q - y_\tau \hat E \hat H_d\!\cdot\! \hat L\ ,
\end{equation}
and
\begin{equation}\label{spH}
\mathcal{W}_{H}= \mu\, \hat H_u\!\cdot\! \hat H_d.
\end{equation}
Higgs superfields are given by
\begin{equation}
\hat{H}_u= \begin{pmatrix}
      \hat H_u^+  \cr
       \hat H^0_u
       \end{pmatrix},\qquad \hat{H}_d= \begin{pmatrix}
      \hat H_d^0  \cr
       \hat H^-_d
       \end{pmatrix}.
 \end{equation}
The extraction of the potential from the superpotential is straightforward and we have $V=V_F+V_D+V_{soft}$ where
\begin{equation}
V_F=\sum_k F_k F_k^*,\quad F_k=\left.\frac{\partial\mathcal W^\dagger}{\partial \hat\Phi_k^\dagger}\right|_{\theta=\bar\theta=0}
\end{equation}
\begin{equation}
 V_D=\frac{1}{2}\sum_k g^2_k ({ \phi^\dagger_i t^a_{ij} \phi_j} )^2 .
 \label{dterm}
 \end{equation}
and $V_{soft}$ is responsible for the soft breaking of supersymmetry. The mass matrices are defined as the second derivative of the potential evaluated on the vacuum. Stop and sbottom mass matrices are given respectively by
\bea\label{stopmssm}
\mathcal M_{\tilde{t}}=\left(
\begin{array}{cc}
m^2_t+m^2_{Q_3}+m_Z^2\cos(2\beta)(\frac{1}{2}-\frac{2}{3}\sin^2\theta_w)\qquad &  m_t(A_t-\mu\cot\beta) \\
   \\
m_t(A_t-\mu\cot\beta) & m_t^2+m^2_{\bar{u}_3}-\frac{1}{3}m_Z^2\cos(2\beta)\sin^2\theta_w
\end{array}
\right)
\eea
\bea\label{sbtmssm}
\mathcal M_{\tilde{b}}=\left(
\begin{array}{cc}
m^2_b+m^2_{Q_3}-m_Z^2\cos(2\beta)(-\frac{1}{2}+\frac{2}{3}\sin^2\theta_w)\qquad &  m_b(A_b-\mu\tan\beta) \\
   \\
m_b(A_b-\mu\tan\beta) & m_b^2+m^2_{\bar{b}_3}+\frac{2}{3}m_Z^2\cos(2\beta)\sin^2\theta_w
\end{array}
\right)
\eea
A close inspection of Eq.~(\ref{stopmssm}) and (\ref{sbtmssm}) shows that the off-diagonal terms are proportional to the quark mass. The mixing $\tilde q_R - \tilde q_L$ is enhanced in the case of stop and hence $\tilde t_1$ is the lightest squark. It's easy to extract the eigenvalues of the matrices in Eq.~(\ref{stopmssm}) and (\ref{sbtmssm}). In the case of stop we have
\bea
m^2_{\tilde t_{1,2}}&=&\frac{1}{2}\Big(m^2_t+m^2_{Q_3}+2m_t^2+\frac{1}{2}m_Z^2\cos(2\beta)\nn\\
&\pm&\sqrt{(m^2_{Q_3}-m^2_{\bar{u}_3}+(\frac{1}{2}-\frac{4}{3}\sin^2\theta_w)\cos(2\beta)m_Z^2)^2+4m_t^2(A_t-\mu\cot\beta)^2}\Big)\nn
\eea

\section{Squark Mass Matrices in TNMSSM}\label{tnmssm}
In this section we give the expression for the mass matrices of stop and sbottom in the model considered in the analysis which is an extension of the MSSM. We have enlarged the Higgs sector of MSSM including a singlet superfield and an SU(2) triplet superfield. The gauge group is the usual SU(3)$_c\times$ SU(2)$_L\times$ U(1)$_Y$. A detailed analysis of this model, named TNMSSM, can be found in \cite{ACPB1,ACPB2,ACPB3}. The superpotential of the model is $\mathcal{W}_{TNMSSM}= \mathcal{W}_{MSSM} + \mathcal{W}_{TS}$ with
\begin{equation}
\mathcal{W}_{MSSM}= y_t \hat U \hat H_u\!\cdot\! \hat Q - y_b \hat D \hat H_d\!\cdot\! \hat Q - y_\tau \hat E \hat H_d\!\cdot\! \hat L\ ,
\label{spm}
 \end{equation}
and
\begin{equation}
\mathcal{W}_{TS}= \lambda_T  \hat H_d \cdot \hat T  \hat H_u\, + \, \lambda_S \hat S  \hat H_d \cdot  \hat H_u\,+ \frac{\kappa}{3}\hat S^3\,+\,\lambda_{TS} \hat S  \textrm{tr}[\hat T^2].
\label{spt}
 \end{equation}
Triplet and doublets superfields are given by
\begin{equation}\label{spf}
 \hat T = \begin{pmatrix}
       \sqrt{\frac{1}{2}}\hat T^0 & \hat T_2^+ \cr
      \hat T_1^- & -\sqrt{\frac{1}{2}}\hat T^0
       \end{pmatrix},\qquad \hat{H}_u= \begin{pmatrix}
      \hat H_u^+  \cr
       \hat H^0_u
       \end{pmatrix},\qquad \hat{H}_d= \begin{pmatrix}
      \hat H_d^0  \cr
       \hat H^-_d
       \end{pmatrix}.
 \end{equation}
In this model the mass matrix for stop is
\bea\label{stop}
\mathcal M_{\tilde{t}}=\left(
\begin{array}{cc}
m^2_t+m^2_{Q_3}+\frac{1}{24} \left(g_Y^2-3g_L^2\right) \left(v_u^2-v_d^2\right)\qquad &  \frac{1}{\sqrt{2}}
   A_T v_u+\frac{y_t v_d}{2} \left( \frac{v_T \lambda _T}{\sqrt 2}- v_S \lambda _S\right) \\
   \\
\frac{1}{\sqrt{2}}
   A_T v_u+\frac{y_t v_d}{2} \left( \frac{v_T \lambda _T}{\sqrt 2}- v_S \lambda _S\right) & m_t^2+m^2_{\bar{u}_3}+\frac{1}{6} \left(v_d^2-v_u^2\right) g_Y^2
\end{array}
\right)
\eea
Comparing Eq.~(\ref{stopmssm}) with Eq.~(\ref{stop}) we can see that they are similar, the only difference being in the off-diagonal terms. The $\mu$-term of Eq.~(\ref{stopmssm}) is now dynamically generated through the terms $ \frac{v_T \lambda _T}{\sqrt 2}- v_S \lambda _S$. This is clear from a comparison of the Higgs part of the superpotential in Eq.~(\ref{spH}) and Eq.~(\ref{spt}).
\section{Branching Ratios of the Stop}\label{branch}
In this section we present the main result of our analysis on the decay of the lightest stop. We are going to compare one of the latest result form the ATLAS collaboration on the mass bound for the lightets stop \cite{stoprecent} with our phenomenological analysis. 

\begin{figure}[bht]
\begin{center}
\mbox{\subfigure[]{\includegraphics[width=.4\linewidth]{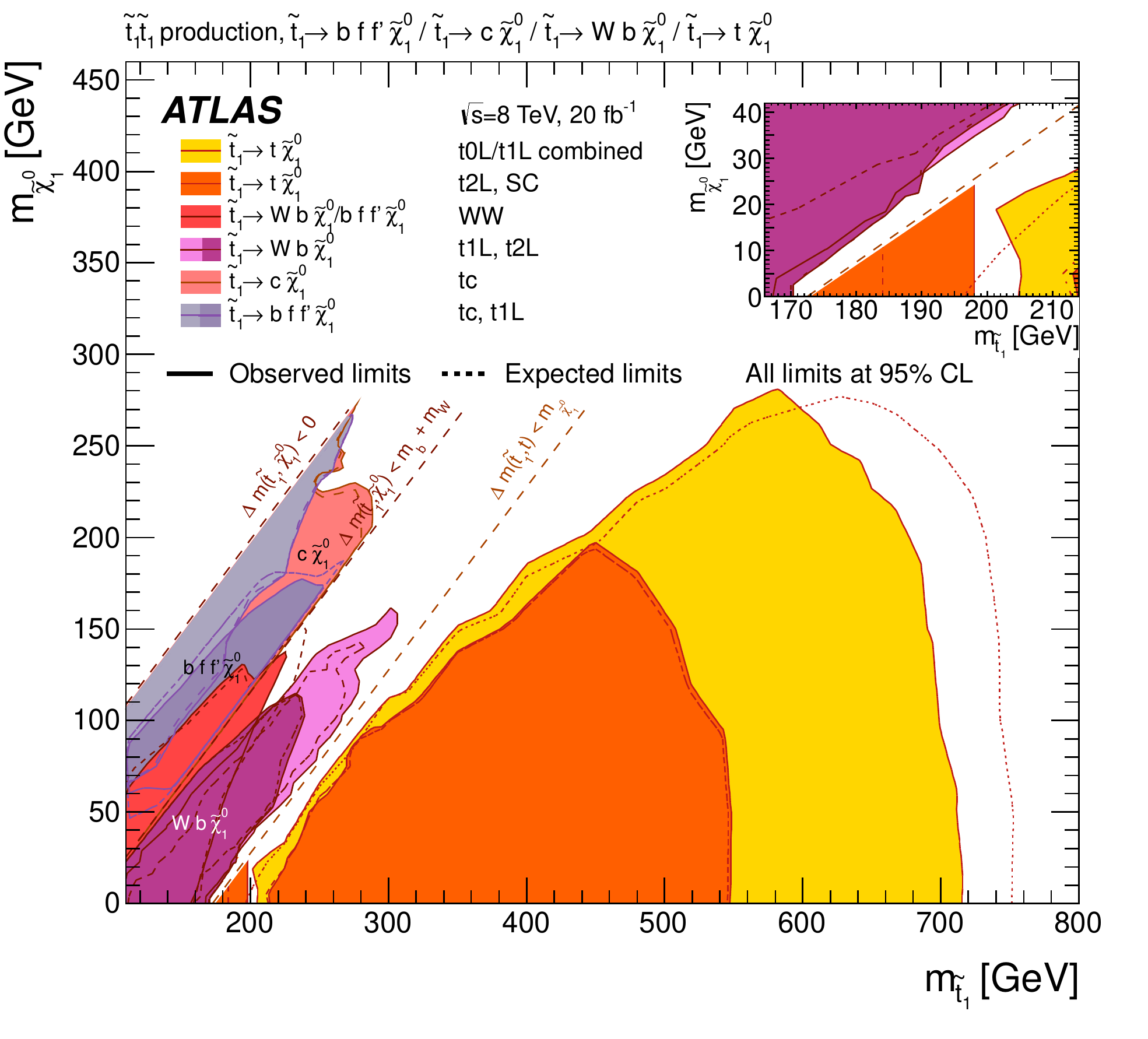}}\subfigure[]{\includegraphics[width=.6\linewidth]{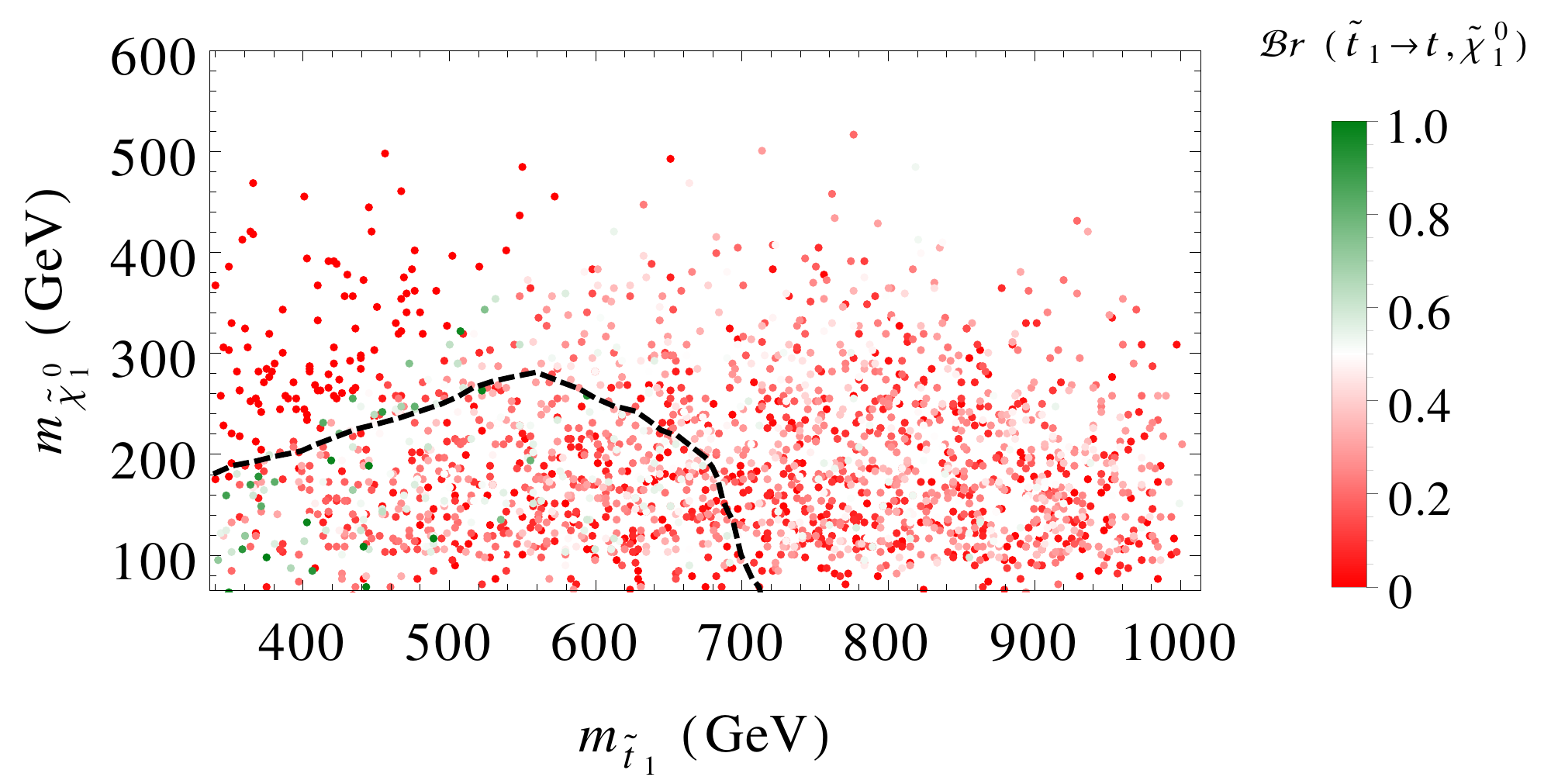}}}
\caption{Plot of the mass correlation $m_{\tilde{t}_1}-m_{\tilde{\chi}_1^0}$ from the ATLAS collaboration \cite{stoprecent} (a) and from our phenomenological analysis (b), where we have reported the correspinding branching ratio for $\tilde{t}_1\to t\,\tilde{\chi}^0_1$. The dashed line is the digitized curve delimiting the yellow region in (a).}\label{figstop}
\end{center}
\end{figure}

In Fig.~\ref{figstop}(a) we can see the correlation plot $m_{\tilde t_1}-m_{\chi_1^0}$  of a recent analysis from the ATLAS collaboration \cite{stoprecent}. In Fig.~\ref{figstop}(b) we present the same correlation plot for a set of $\sim2000$ benchmark points used in a previous analysis. The dashed line is the digitized curve delimiting the yellow region of Fig.~\ref{figstop}(a).
Let us remark that the benchmark points showed in Fig.~\ref{figstop}(b) pass a certain number of phenomenological tests. In particular for each point we have a Higgs boson with a mass $\sim125$ GeV and the couplings with the Z bosons, the W bosons, the photons and the gluons are within the current experimental limits \cite{ACPB3}. Moreover the mass of the lightest neutralino, $\chi_1^0$, which is the lightest supersymmetric particle (LSP) of the model, is compatible with the bounds coming from the relic abundance of dark matter.

The bounds on the mass of the lightest stop are putted under the assumption of $\mathcal{B}\textit{r}\,(\tilde{t}_1\to t\,\tilde{\chi}^0_1)=1$ \cite{stoprecent}. This is not the case if the model has an extended Higgs sector, as one can see from Fig.~\ref{figstop}(b).
Another possible decay channel for the stop is $\tilde{t}_1\to b\, \tilde{\chi}_1^\pm$ and the mass bounds for the stop mass coming from this analysis are presented in \cite{bcha}. The channel $\tilde{t}_1\to b\, \tilde{\chi}_1^\pm$ is assumed to have $\mathcal{B}\textit{r}=1$. 
\begin{figure}[ht]
\begin{center}
\mbox{\hspace{-.5cm}\subfigure[]{\includegraphics[width=.48\linewidth]{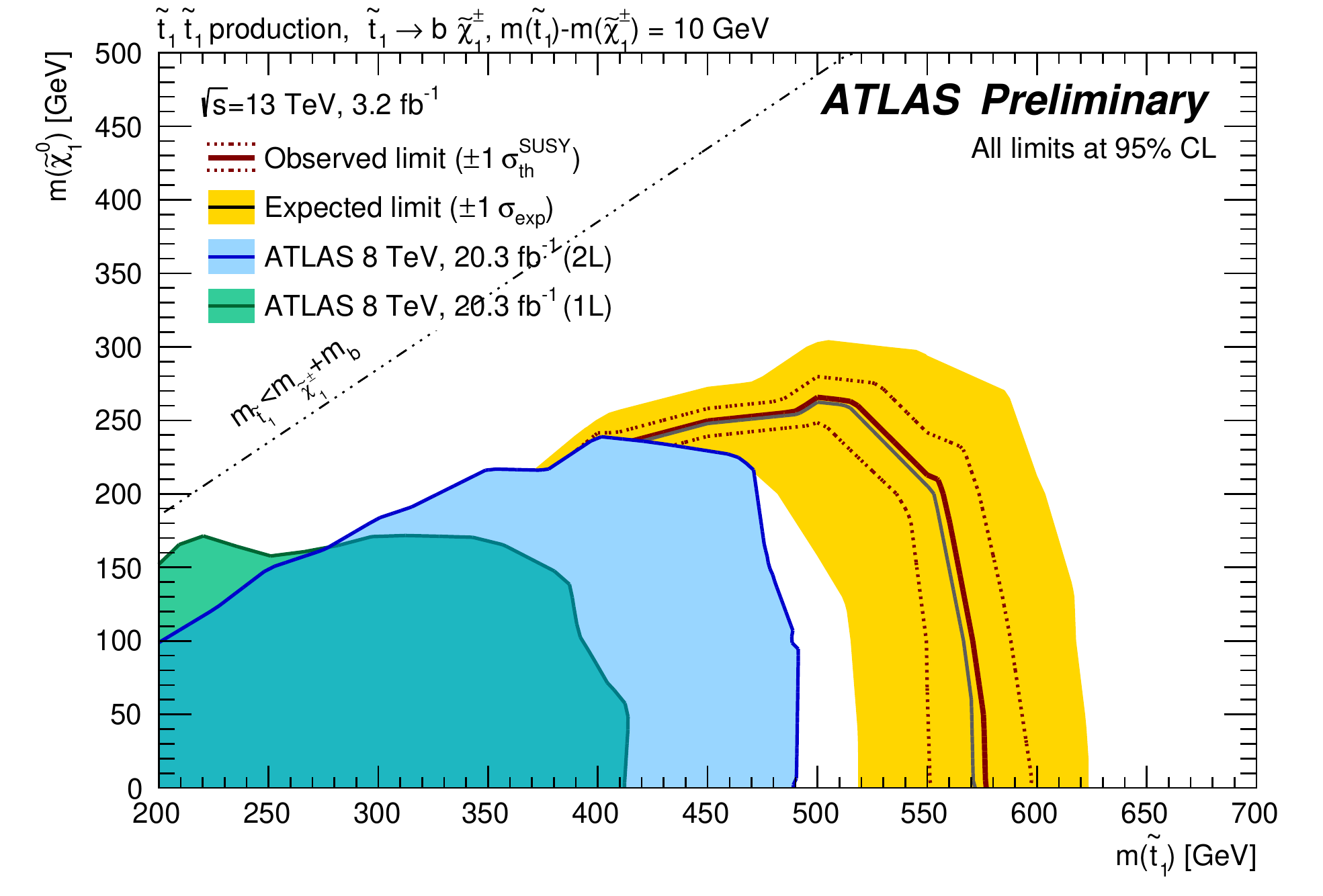}}\subfigure[]{\includegraphics[width=.6\linewidth]{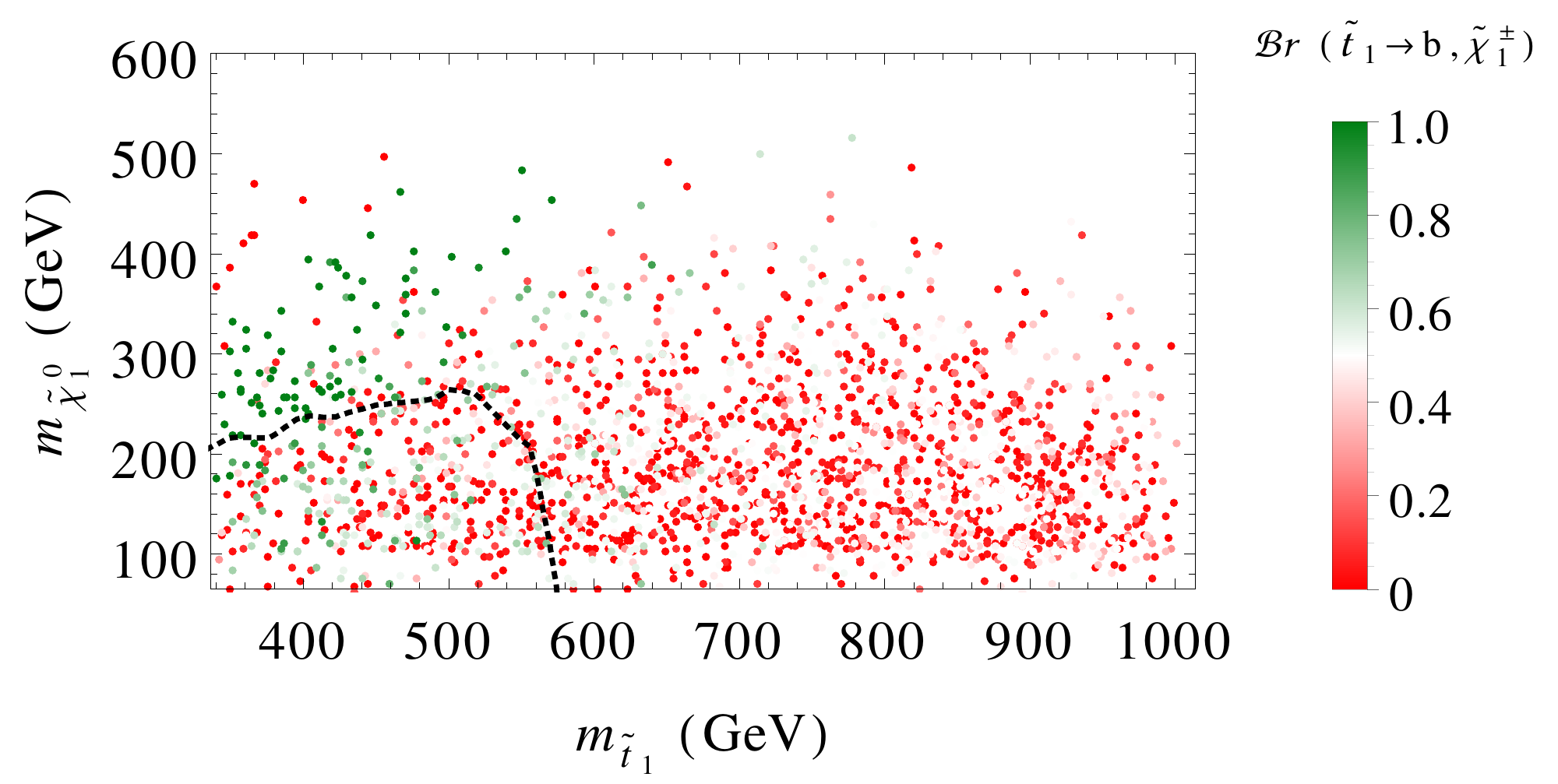}}}
\caption{Plot of the mass correlation $m_{\tilde{t}_1}-m_{\tilde{\chi}_1^0}$ from the ATLAS collaboration \cite{bcha} (a) and from our phenomenological analysis (b), where we have reported the correspinding branching ratio for $\tilde{t}_1\to b\,\tilde{\chi}^\pm_1$. The dashed line is the digitized curve corresponding to the observed limit in (a).}\label{cha}
\end{center}
\end{figure}
In Fig.~\ref{cha}(a) is shown the correlation plot $m_{\tilde t_1}-m_{\chi_1^0}$ in the case of  $\tilde{t}_1\to b\, \tilde{\chi}_1^\pm$ and in Fig.~\ref{cha}(b) we present the results of our analysis. As is the previous case the dashed line is the digitized curve corresponding to the observed limit in Fig.~\ref{cha}(a). In this case there are some points with $\mathcal{B}\textit{r}_{b\, \tilde{\chi}_1^\pm}\simeq1$ but they are outside the excluded region. The points below the dashed curve have $\mathcal{B}\textit{r}_{b\, \tilde{\chi}_1^\pm} <1$.

We clearly see that the mass bounds for the scalar top considered here are evaded by a large number of benchmark points. In fact in both Fig.~\ref{figstop}(b) and Fig.~\ref{cha}(b) there are mostly red points corresponding to branchings lower than 50\%. 

\section{Conclusions}\label{concl}
We have presented a phenomenological analysis of the decay channels of the scalar top in the context of a supersymmetric theory with an extended Higgs sector. We have compared our results with the recent analysis done by the ATLAS collaboration \cite{stoprecent,bcha} for the decay channels $\tilde{t}_1\to t\,\tilde{\chi}^0_1$ and $\tilde{t}_1\to b\, \tilde{\chi}_1^\pm$. In both the cases the branching ratios for these channels is assumed to be 100\%. This is not true in the case of a non-minimal SUSY theory, like the TNMSSM \cite{ACPB1,ACPB2,ACPB3}, as shown in Fig.~\ref{figstop}(b) and Fig.~\ref{cha}(b). Although the mass matrices of the squarks are almost the same in the MSSM and in the TNMSSM, the branching ratios of the channels studied at the LHC are quite different. This is due to the enlarged Higgs sector of the TNMSSM and to the presence of other representation of SU(2)$_L$ than the doublets, as extensively discussed in \cite{ACPB1,ACPB2,ACPB3}.

\end{document}